\pdfoutput=1
\documentclass{raa}

\usepackage{mathptmx}
\usepackage{ae,aecompl}

\usepackage{graphicx}	% Including figure files
\usepackage{amsmath}	% Advanced maths commands
\usepackage{amssymb}	% Extra maths symbols
\usepackage{pifont}
\usepackage{txfonts}
\usepackage{url}
\usepackage{subfigure}
\usepackage{longtable}
\usepackage{lscape}
\usepackage{multirow}
\usepackage{color}
\usepackage{textcomp}
\usepackage{natbib}
\usepackage[colorlinks,linkcolor=red,anchorcolor=green,citecolor=blue]{hyperref}

\newcommand{\kms}{{\rm km~s}^{-1}}
\newcommand{\HII}{H\,{\small{II}} }
\newcommand{\CII}{C\,{\small{II}} }

\newcommand{\dotsec}{\rlap.{''}}

\begin{document}

\title{Radio Recombination Line Observations at $1.0 - 1.5$\,GHz with FAST}
%\footnotetext{$*$ Supported by the National Natural Science Foundation of China.}

%   \subtitle{I. Place Your Subtitle Here}

   \volnopage{Vol.0 (2021) No.0, 000--000}      %%preserved for Editor. DOn't remove!
   \setcounter{page}{1}          %%starting page, preserved for Editor. DOn't remove!
   \author{Chuan-Peng Zhang\inst{1},
      Jin-Long Xu\inst{1},
      Guang-Xing Li\inst{2},
      Li-Gang Hou\inst{1},
      Nai-Ping Yu\inst{1},
      Peng Jiang\inst{1}
   }

   \institute{National Astronomical Observatories, Chinese Academy of Sciences, 100101 Beijing, China; {\it cpzhang@nao.cas.cn} \\
    \and
    South-Western Institute for Astronomy Research, Yunnan University, Kunming, 650500 Yunnan, P.R. China\\
   }

   %\date{Received~~2016 month day; accepted~~2016~~month day}

\abstract{\HII regions made of gas ionized by radiations from young massive stars,  are widely distributed in the Milky Way. They are tracers for star formation, and their distributions are correlated with the Galactic spiral structure. Radio recombination lines (RRLs) of hydrogen and other atoms allows for the precisest determination of physical parameters such as temperature and density. However, RRLs at around 1.4\,GHz from \HII regions are weak and their detections are difficult. As a result, only a limited number of detections have been obtained yet. The 19-beam receiver on board of the Five-hundred-meter Aperture Spherical radio Telescope (FAST) can simultaneously cover 23 RRLs for H$n\alpha$, He$n\alpha$, and C$n\alpha$ ($n=164-186$), respectively. This, combined with its unparalleled collecting area, makes FAST the most powerful telescope to detect weak RRLs. In this pilot survey, we use FAST to observe nine \HII regions at L band. We allocate 20\,minutes pointing time for each source to achieve a sensitivity of around 9\,mK in a velocity resolution of 2.0\,$\kms$. In total, 21 RRLs for H$n\alpha$ and C$n\alpha$ at $1.0-1.5$\,GHz have been simultaneously detected with strong emission signals. Overall, the detection rates for the H$167\alpha$ and C$167\alpha$ RRLs are 100\%, while that for the He$167\alpha$ RRL is 33.3\%. Using hydrogen and helium RRLs, we measure the electron density, electron temperature, and pressure for three \HII regions. This pilot survey demonstrates the capability of FAST in RRL measurements, and a statistically meaningful sample with RRL detection, through which knowledges about Galactic spiral structure and evolution can be obtained, is expected in the future.
 }

   \authorrunning{Chuan-Peng Zhang, et al.}                     % author_head in even pages
   \titlerunning{RRL Observations with FAST}       % title_head in odd pages

   \maketitle

%%________________________________________________________________

\section{Introduction}    %% first-level sections will be auto-capitalized
\label{sect:intro}

Galactic \HII regions are made of gas ionized young massive stars. They are tracers for star formation, their evolutions are interesting from an astrophysical point of view, and their distributions are correlated with the Galactic spiral structure \citep[e.g.,][]{Churchwell2006,Churchwell2007,Beaumont2010,Hou2014,Zhang2014,Zhang2017}. \HII regions are ubiquitously distributed in the Milky Way. The Wide-Field Infrared Survey Explorer (WISE) catalog of \HII regions \citep{Anderson2014} is the largest sample of Galactic \HII region candidates to date. The sample are distributed throughout the Milk Way disk, making them idea tracers for the Galactic spiral structure. Determining their distances and physical parameters is thus crucial.

Radio recombination lines (RRLs) of hydrogen provide the simplest and precisest method of determining the electron temperature and density of \HII regions. Unlike observations of optical emission lines, RRLs are unaffected by reddening from interstellar dust. Further, they can be accurately measured even in weak astronomical sources \citep{Gordon2002}, allowing for accurate determination of the physical parameters. RRLs are distributed in a wide range of frequencies. Recent RRL surveys at different wavelengths were performed with highly sensitive facilities to study \HII regions \citep[e.g.,][]{Anderson2011,Anderson2014,Liu2013,Alves2015,Chen2020}. However, the RRLs at around 1.4\,GHz from \HII regions are relatively weak \citep{Anderson2011}, thus either large telescopes or long integration are needed to detect them. Before FAST, these observations are difficult, and only a small sample of \HII regions were detected at $\sim$1.4 GHz.

FAST is located at geographic latitude of $25^\circ39'10\dotsec6$ and its observable maximum zenith angle is $40^\circ$ \citep{Nan2011,Jiang2019,Jiang2020}, thus FAST observation can cover all the \HII regions distributed in the Galactic longitude range of around 21$^\circ$ to 223$^\circ$. The 19-beam receiver of the FAST has a band width of 500\,MHz with a frequency range of $1.0-1.5$\,GHz \citep{Jiang2019,Jiang2020}, which can simultaneously cover 23 RRLs for H$n\alpha$, C$n\alpha$, and He$n\alpha$ (principal quantum number $n=164-186$), respectively. This makes FAST the most powerful telescope to detect weak RRLs at L band. Therefore, a larger population of \HII regions deserve to be systematically studied with FAST. Using the RRLs and continuum emission at around 1.4\,GHz, we could measure the pressure, evolutionary ages, distances, and relative abundance of elements in \HII regions. With a reasonably-sized sample, we could determine the distribution of them in the Milky Way disk and correlate it with the physical parameters of the \HII regions. This will deepen our understanding of the co-evolution between \HII regions and the Milky Way.

In this work, as a pilot study, nine \HII region candidates are selected from a $4-6$\,GHz RRL survey with Shanghai Tianma 65\,m telescope by \citet{Chen2020}, where RRLs such as H$n\alpha$, He$n\alpha$, and C$n\alpha$ ($n=98-113$) have been detected (see Table\,\ref{tab_source}). We observe $1.0 - 1.5$\,GHz (see Table\,\ref{tab_frequency}) RRLs using FAST to determine the physical parameters and to provide a first constraint on the expected detection rate.

\begin{table*}
%\begin{minipage}[t]{\columnwidth}
\caption{Sources and observational results.}
\label{tab_source} 
\centering \small  %\footnotesize %\scriptsize
\setlength{\tabcolsep}{1.6mm}{
\begin{tabular}{lccccccccc}
\hline \hline
Source &  R.A.(J2000)&  Dec.(J2000) &  Velocity & $T_{\rm A}$  & FWHM & $T_{\rm A}$  & FWHM  & $T_{\rm A}$  & FWHM  \\
      &   &   &  & \multicolumn{2}{c}{(H167$\alpha$)} & \multicolumn{2}{c}{(C167$\alpha$)} & \multicolumn{2}{c}{(He167$\alpha$)} \\
       &  hh:mm:ss   & dd:mm:ss  & $\kms$  &   K & $\kms$ &   K & $\kms$ &   K & $\kms$  \\  
\hline
G43.148$+$0.013  &  19:10:10.98  &  $+$09:05:18.2  &  10.47  &  0.244  &  28.26  &  0.074  &  18.07  \\        
G43.177$-$0.008  &  19:10:18.73  &  $+$09:06:13.4  &  15.07  &  0.242  &  29.04  &  0.066  &  16.64  \\        
G48.905$-$0.261  &  19:22:07.85  &  +14:03:12.6  &  60.78  &  0.203  &  27.71  &  0.028  &  34.13  \\        
G48.946$-$0.331  &  19:22:27.89  &  +14:03:24.9  &  65.85  &  0.251  &  26.53  &  0.045  &  14.18  &  0.017  &  6.02  \\
G48.991$-$0.299  &  19:22:26.22  &  +14:06:42.2  &  66.61  &  0.229  &  28.80  &  0.038  &  13.07  &  0.020  &  5.55  \\
G49.028$-$0.217  &  19:22:12.74  &  +14:10:58.4  &  61.91  &  0.295  &  21.26  &  0.047  &  11.93  &  0.017  &  5.07  \\
G49.224$-$0.334  &  19:23:01.08  &  +14:18:00.6  &  63.43  &  0.154  &  30.58  &  0.037  &  31.37  \\       
G49.368$-$0.303  &  19:23:11.17  &  +14:26:32.6  &  49.47  &  0.226  &  29.71  &  0.049  &  29.15  \\        
G49.466$-$0.408  &  19:23:45.74  &  +14:28:45.2  &  56.66  &  0.169  &  25.84  &  0.053  &  23.23  \\    
 \hline
\end{tabular}}
\begin{flushleft}
%\begin{tabular}{l}
%\normalsize
\textbf{Notes.} Only the $n=167$ data for H$n\alpha$, He$n\alpha$, and C$n\alpha$ RRLs are listed above, because other transitions have a very close spectral information to the $n=167$. \\
%\end{tabular}
\end{flushleft}
%\end{minipage}
\end{table*}

\section{Observations and Data Reduction}

\subsection{Observations with FAST}

The FAST observations were conducted between Aug.28 and Sep.19, 2020. We used tracking mode to point to each target using the central beam (Beam-M01) of the 19-beam receiver on FAST. Other beams were used as off-source observations to calibrate the data in Beam-M01. This setup could save lots of observation time. The half-power beamwidth (HPBW) is $\sim$2.9$'$ at 1.4\,GHz , and the pointing error is $\sim$0.15$'$ \citep{Jiang2019,Jiang2020}. For the backend, a high spectral resolution (1048576 channels), full bandwidth (500\,MHz), and dual polarization mode were used to collect the observational signal, resulting a frequency resolution of 476\,Hz or a velocity resolution of $\sim$0.1\,$\kms$ at 1.4\,GHz. Each source was integrated for 20\,minutes with a sampling rate of one second. For intensity calibration, noise signal with amplitude of 1\,K was injected under a period of two seconds in only first two minutes at the beginning of each source observation. The relevant degrees per flux unit factor (DPFU) for Beam-M01 is $\rm DPFU\sim16.02\pm0.26$ Kelvin per Jy\,beam$^{-1}$ at 1400\,MHz \citep{Jiang2020}.

\subsection{Data Reduction}

Our selected sources have extremely strong continuum emission ($\gtrsim$10\,Jy), this leads to that the total fluxes with noise diode at switch-on and switch-off states have no obvious discrepancy. Thus, our observational data could not be calibrated directly by noise diode. This issue could be solved by using the simultaneously observed data in other beams to calibrate our target data from the central beam, assuming that all data from each beam have the same rms. We could firstly calibrate data from the FAST Beam-M16 receiver \citep{Jiang2020} to get the antenna temperature ($T_{\rm A}$) using:
\begin{equation}
  T_{\rm A} = T_{\rm noise}\times\frac{\rm Power_{cal\,off}}{\rm Power_{cal\,on} - Power_{cal\,off}},
  \label{equ_calib}
\end{equation}
where $T_{\rm noise}$ is the injected noise temperature of known, ${\rm Power_{cal\,on}}$ and ${\rm Power_{cal\,off}}$ are, respectively, the observed total power for noise diode with switch-on and switch-off. Assuming that the FAST Beam-M01 data have the same rms as the Beam-M16, we then could calibrate the observed data in the Beam-M01 using the Beam-M16\footnote{For a test, we also use other beam data to calibrate the Beam-M01 data, there is no obvious discrepancy in the final results.}. Such calibration method may introduce an uncertainty of 30\% into the derived antenna temperature. Finally, we smoothed the spectral data into a resolution (channel width) of 10\,KHz, corresponding to a spectral resolution of around $2.0\,\kms$ at 1.4\,GHz. The sensitivity achieves to around 9\,mK at such conditions.

\begin{figure*}
\centering
\includegraphics[width=0.99\textwidth, angle=0]{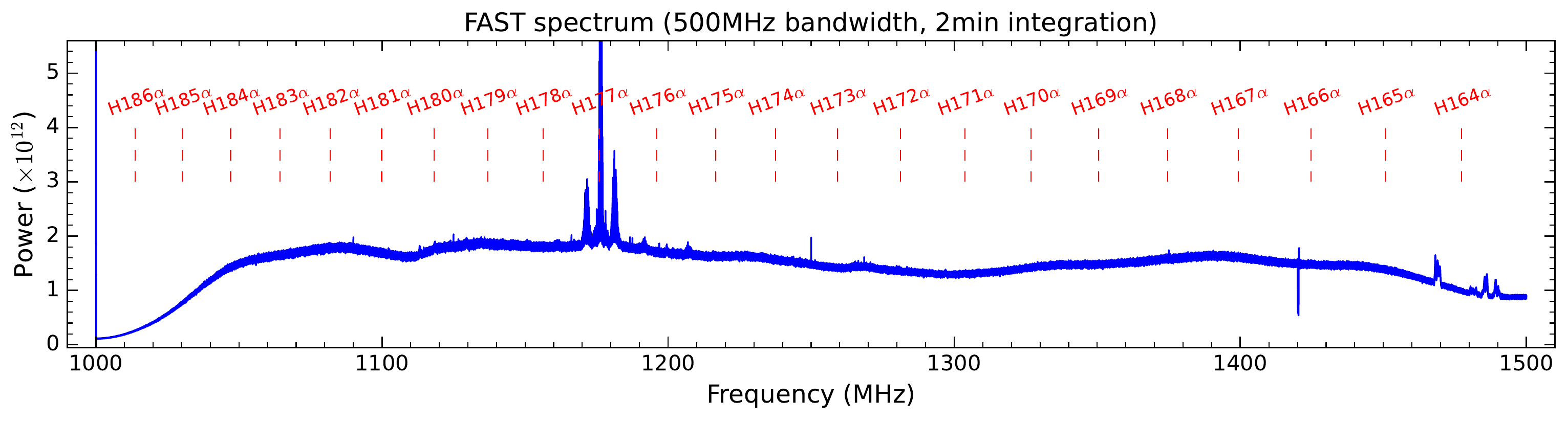}
\caption{H$n\alpha$ ($n=164-186$) RRLs in a FAST spectrum for G43.148$+$0.013. The blue spectrum shows original and uncalibrated data with 500\,MHz bandwidth and two minutes integration from FAST observations.}
\label{Fig:fast_spectrum}
\end{figure*}

\begin{figure*}
\centering
\includegraphics[width=0.32\textwidth, angle=0]{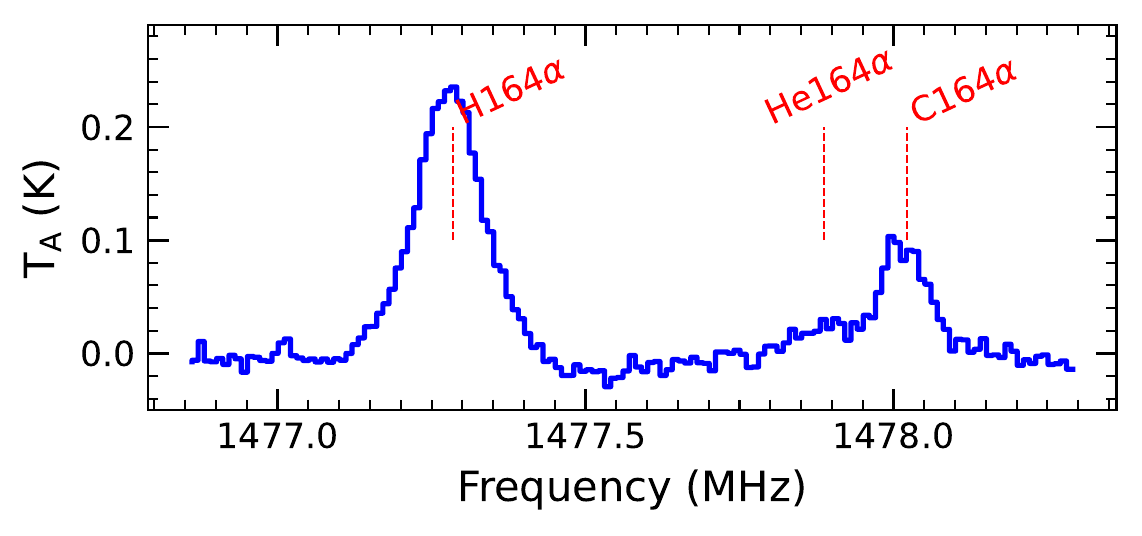}
\includegraphics[width=0.32\textwidth, angle=0]{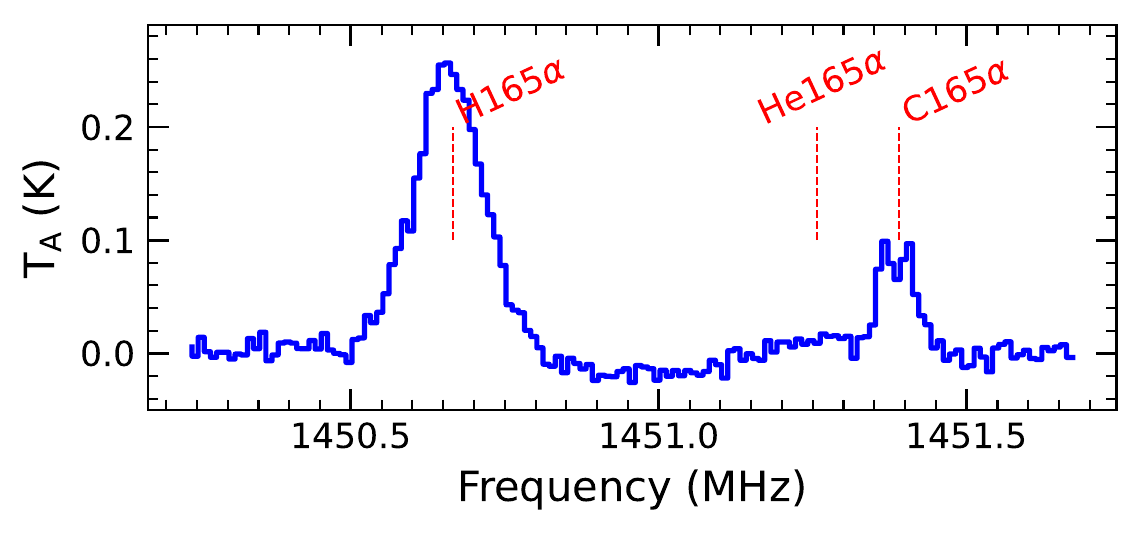}
\includegraphics[width=0.32\textwidth, angle=0]{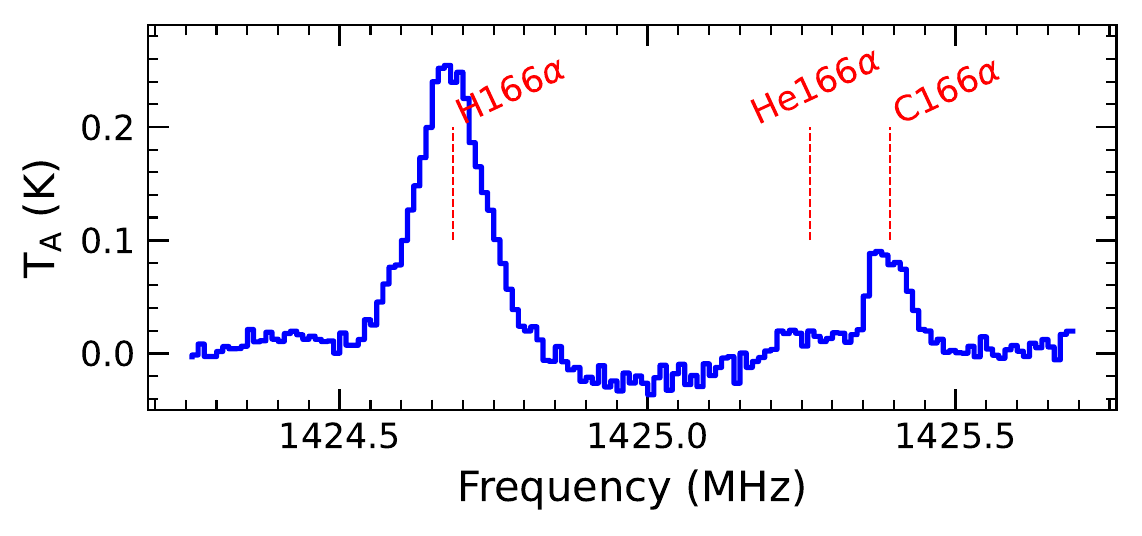}
\includegraphics[width=0.32\textwidth, angle=0]{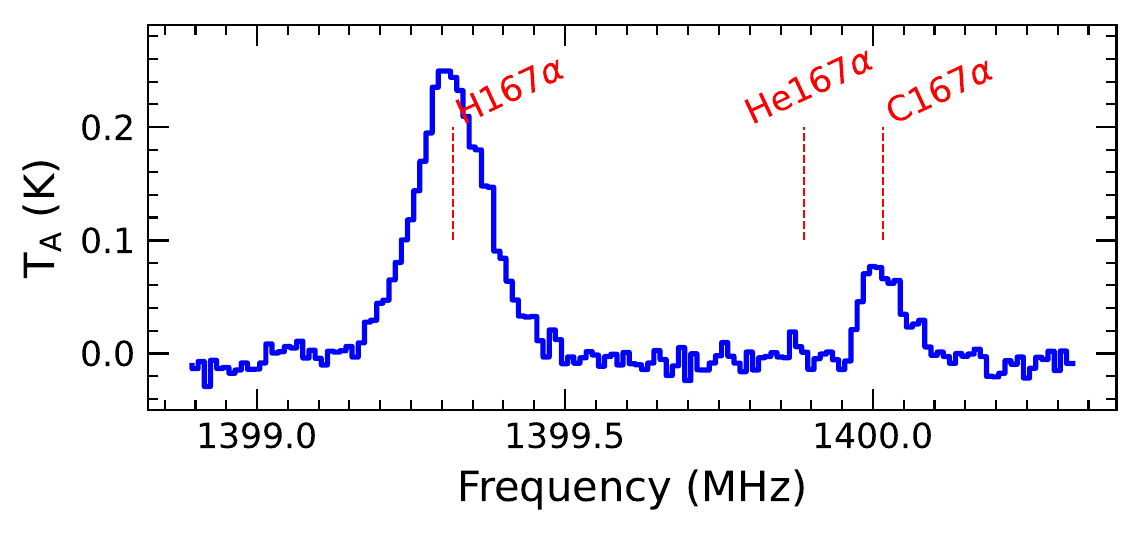}
\includegraphics[width=0.32\textwidth, angle=0]{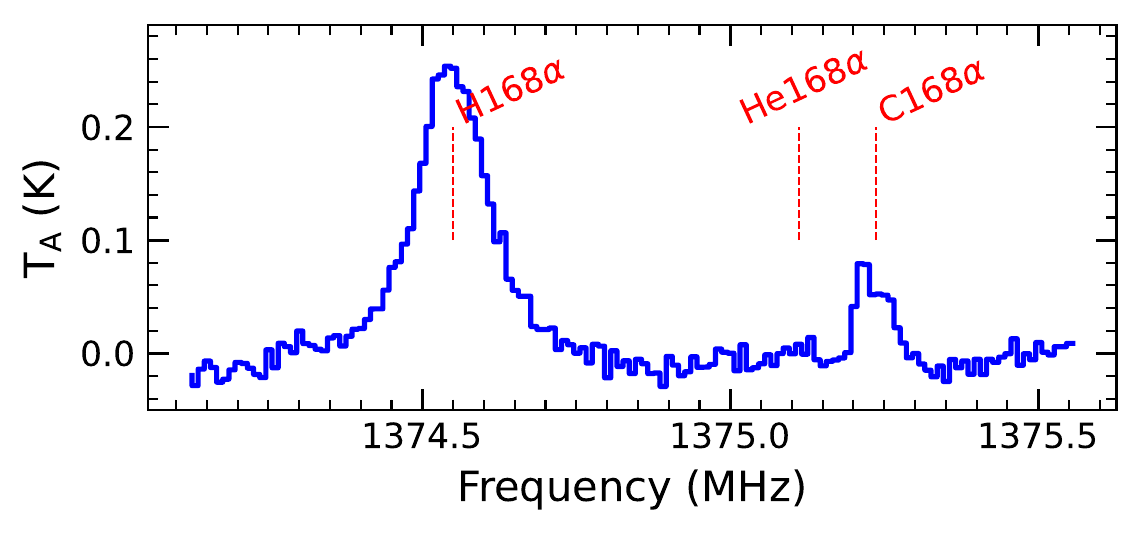}
\includegraphics[width=0.32\textwidth, angle=0]{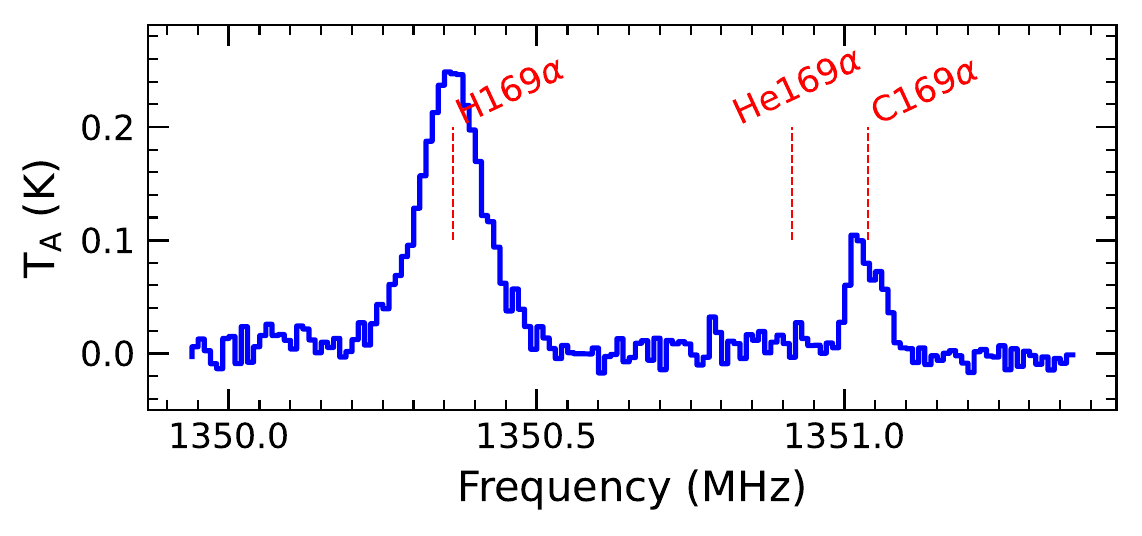}
\includegraphics[width=0.32\textwidth, angle=0]{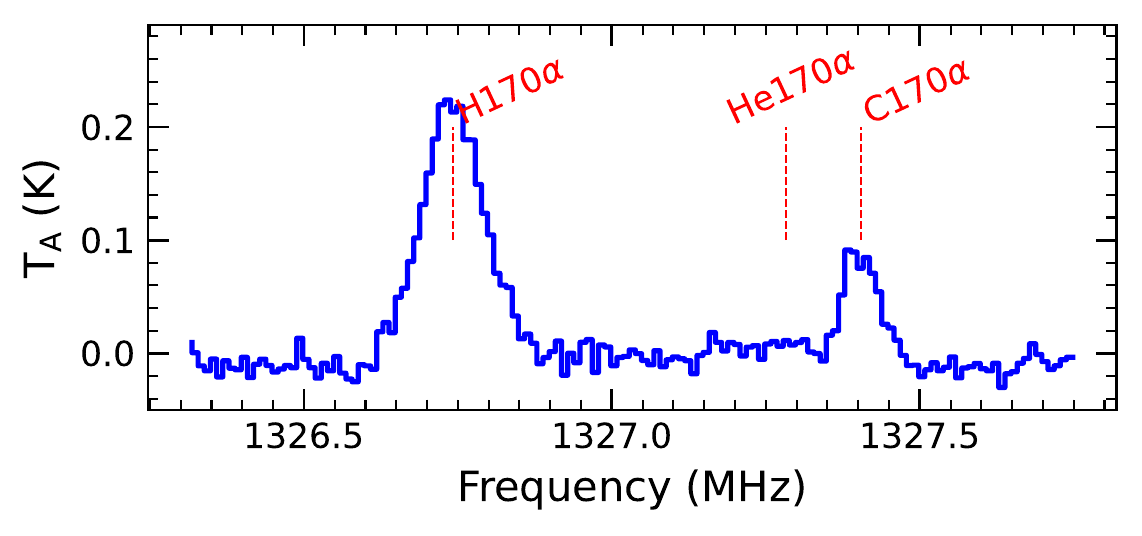}
\includegraphics[width=0.32\textwidth, angle=0]{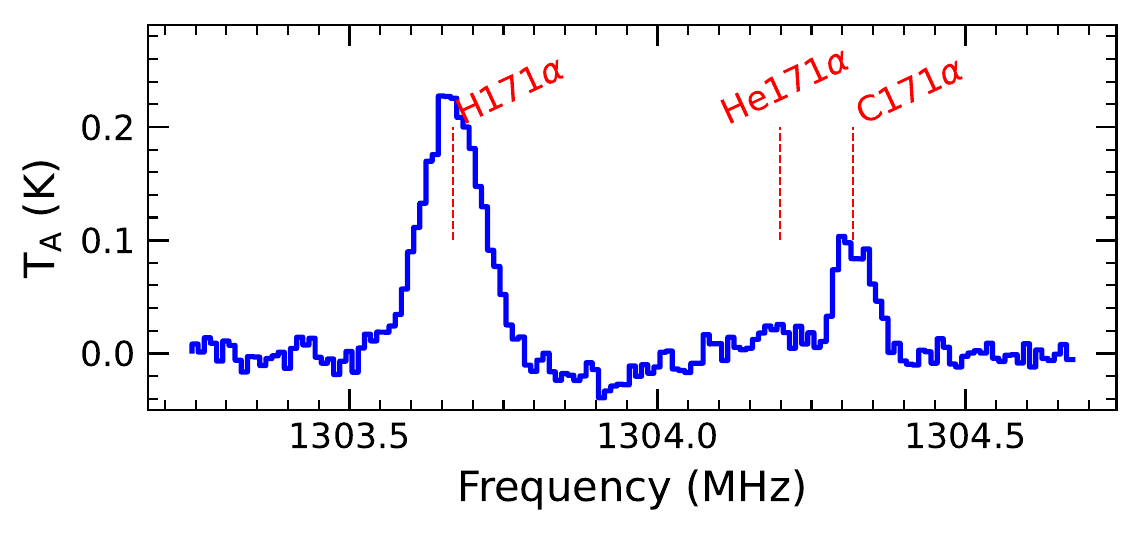}
\includegraphics[width=0.32\textwidth, angle=0]{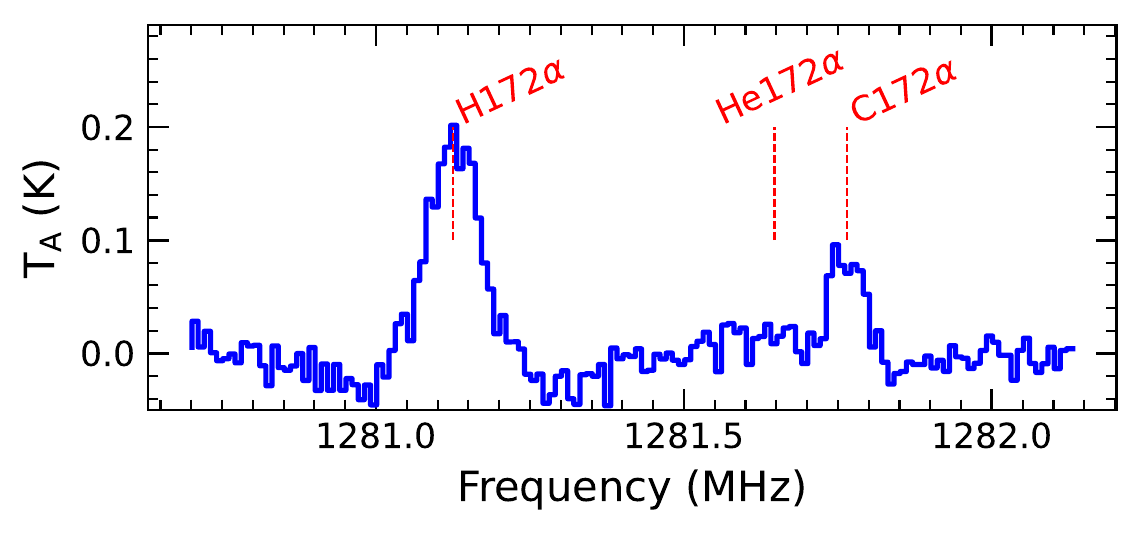}
\includegraphics[width=0.32\textwidth, angle=0]{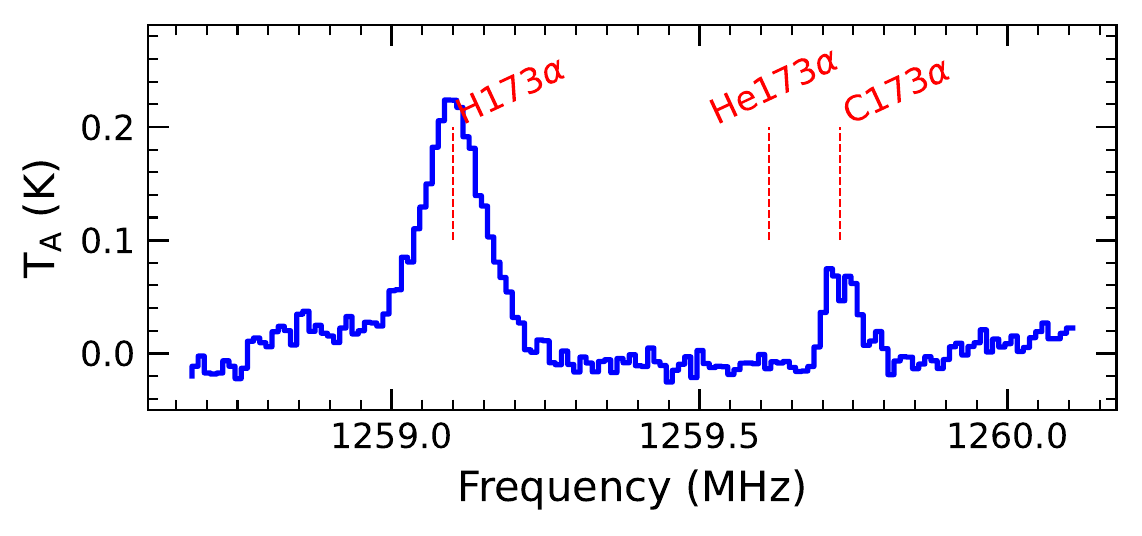}
\includegraphics[width=0.32\textwidth, angle=0]{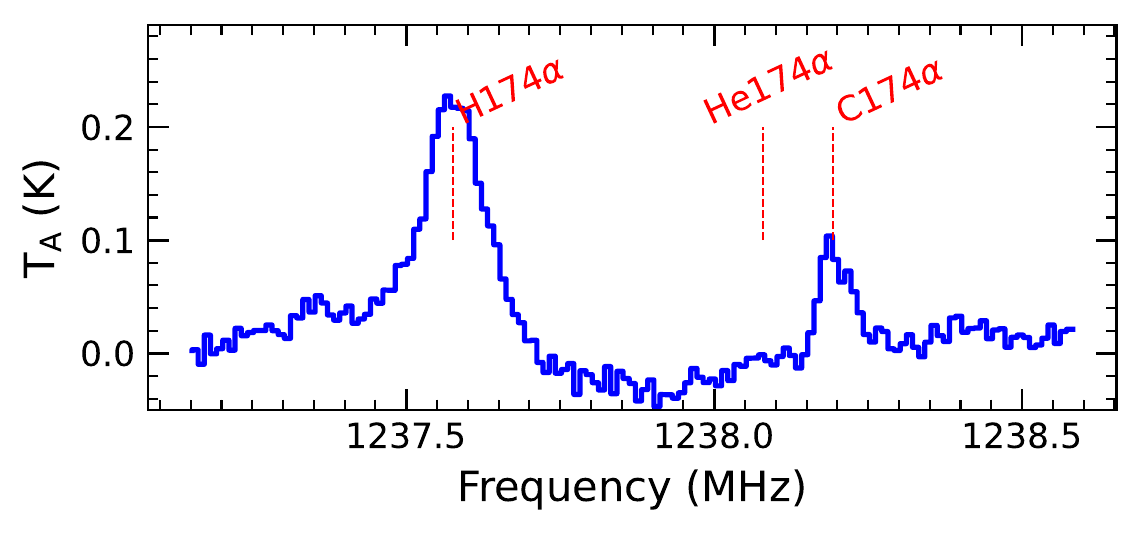}
\includegraphics[width=0.32\textwidth, angle=0]{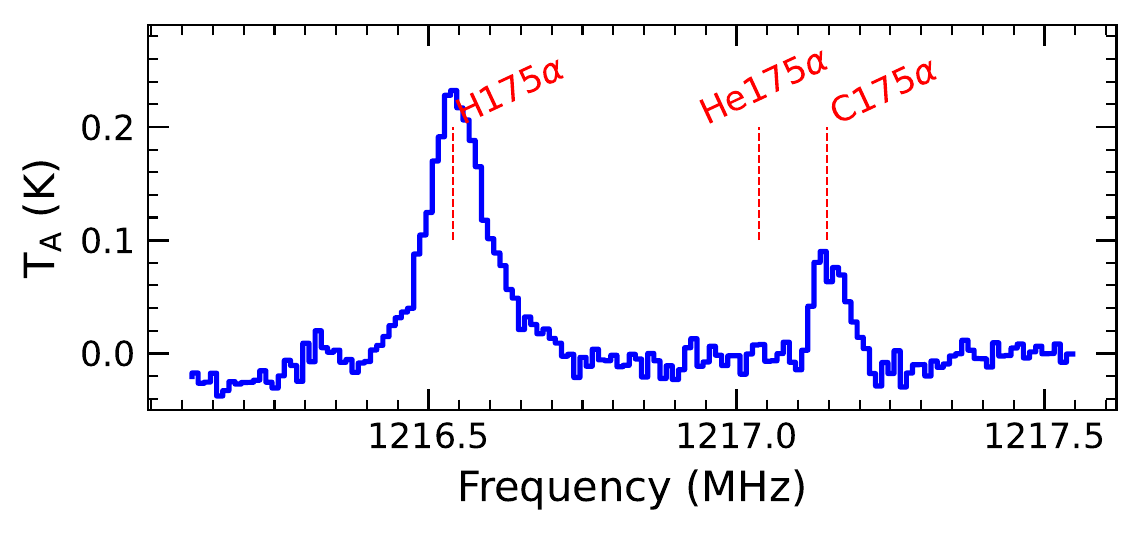}
\includegraphics[width=0.32\textwidth, angle=0]{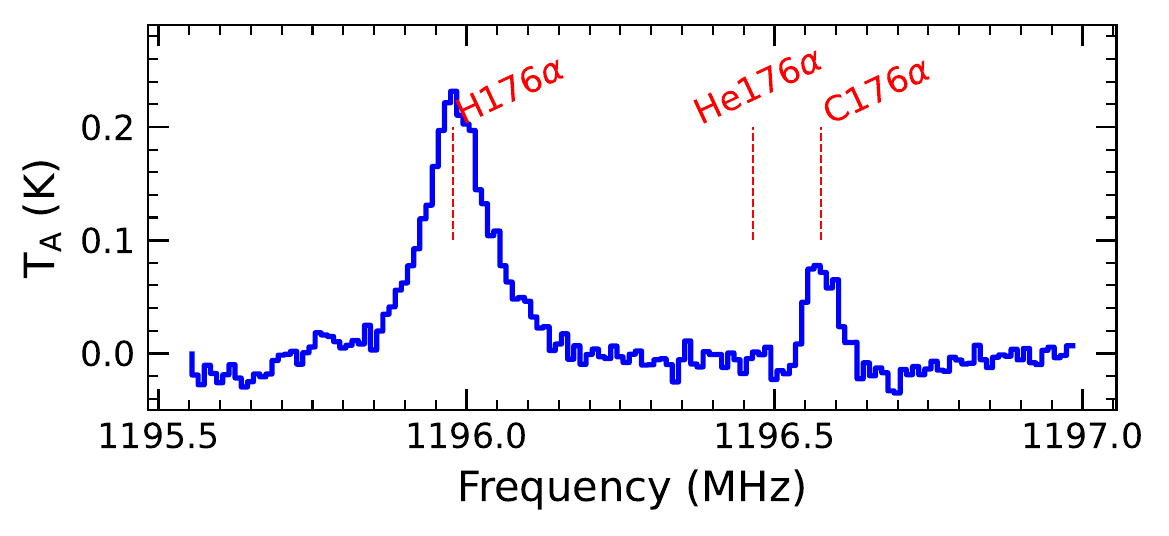}
\includegraphics[width=0.32\textwidth, angle=0]{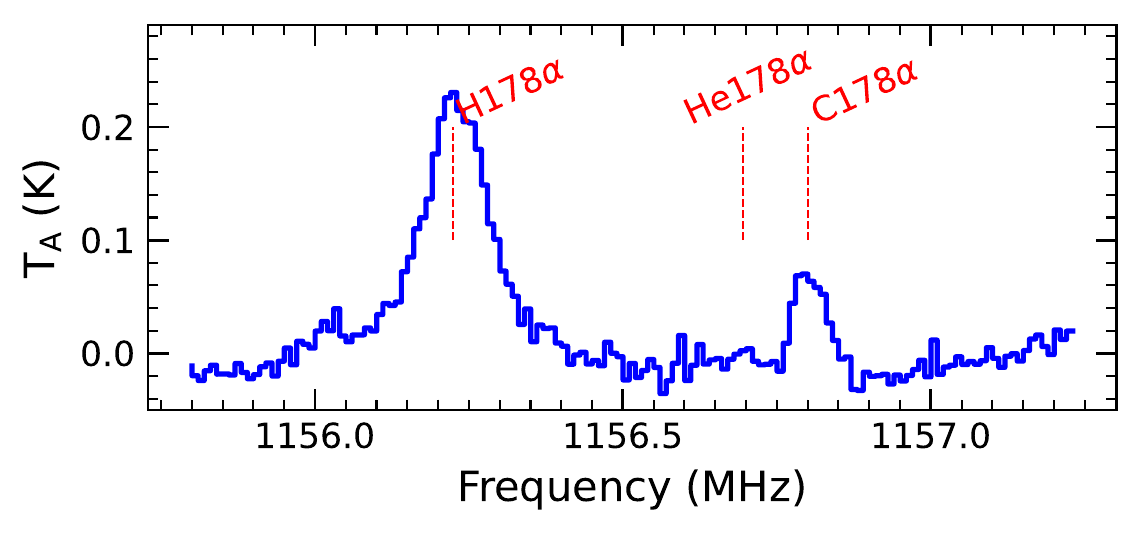}
\includegraphics[width=0.32\textwidth, angle=0]{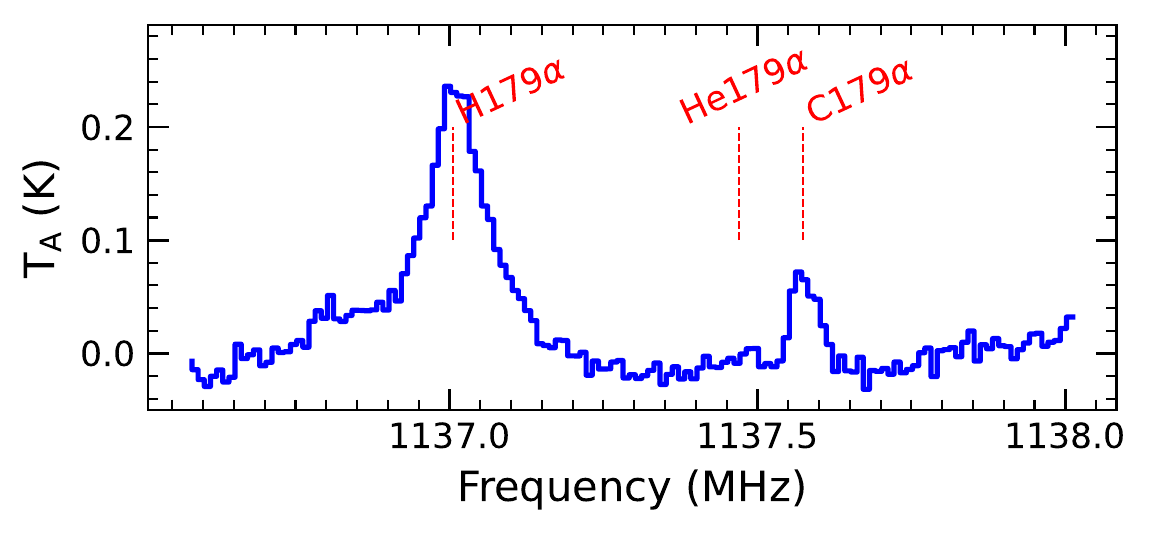}
\includegraphics[width=0.32\textwidth, angle=0]{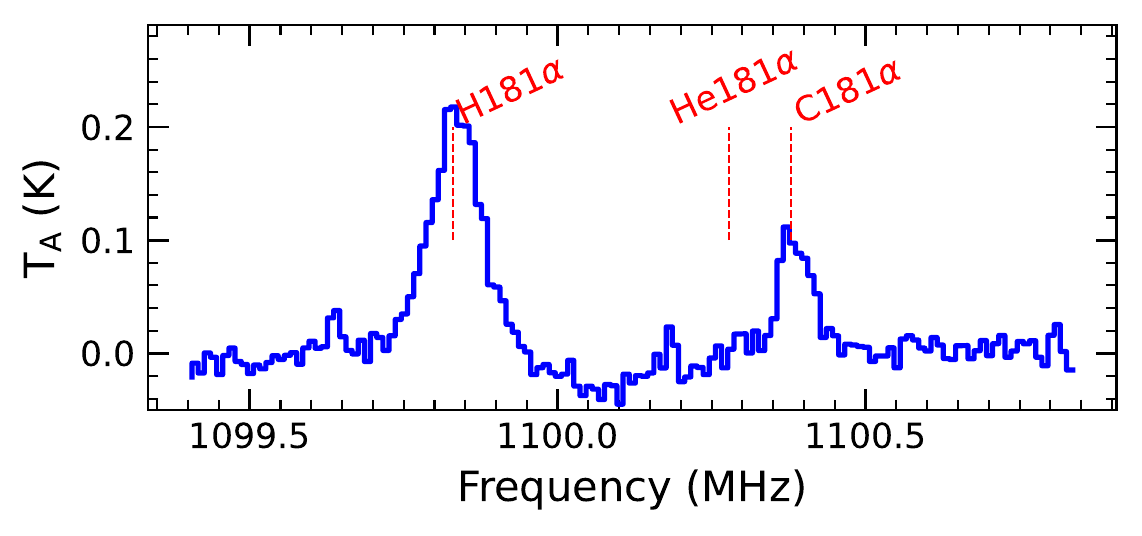}
\includegraphics[width=0.32\textwidth, angle=0]{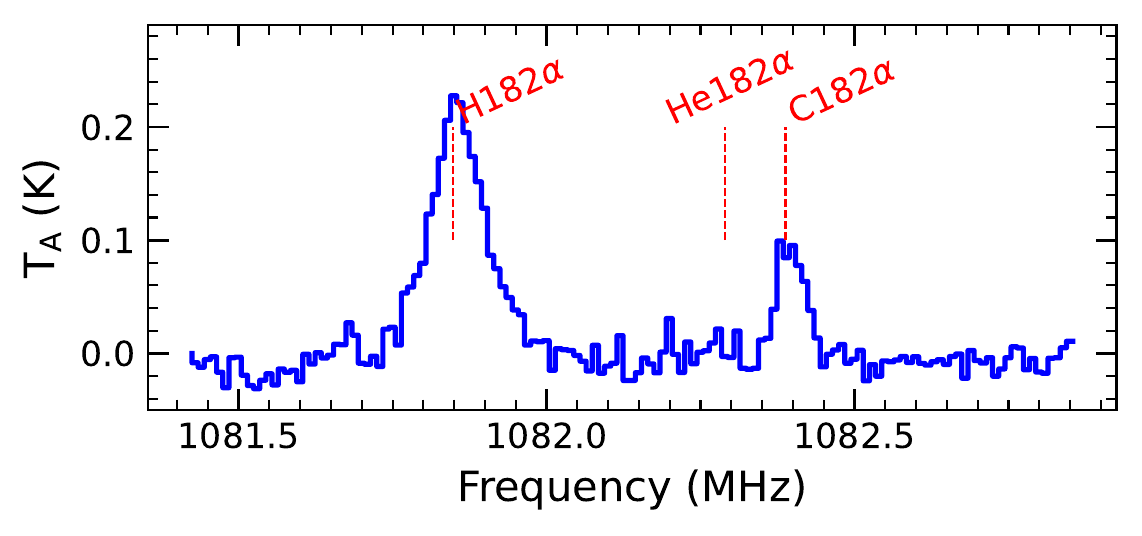}
\includegraphics[width=0.32\textwidth, angle=0]{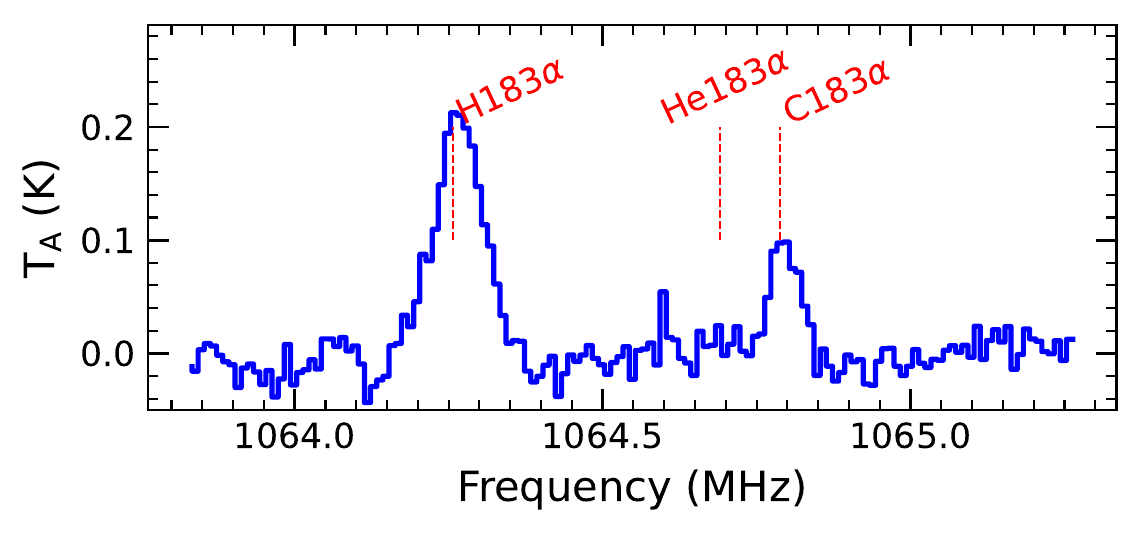}
\includegraphics[width=0.32\textwidth, angle=0]{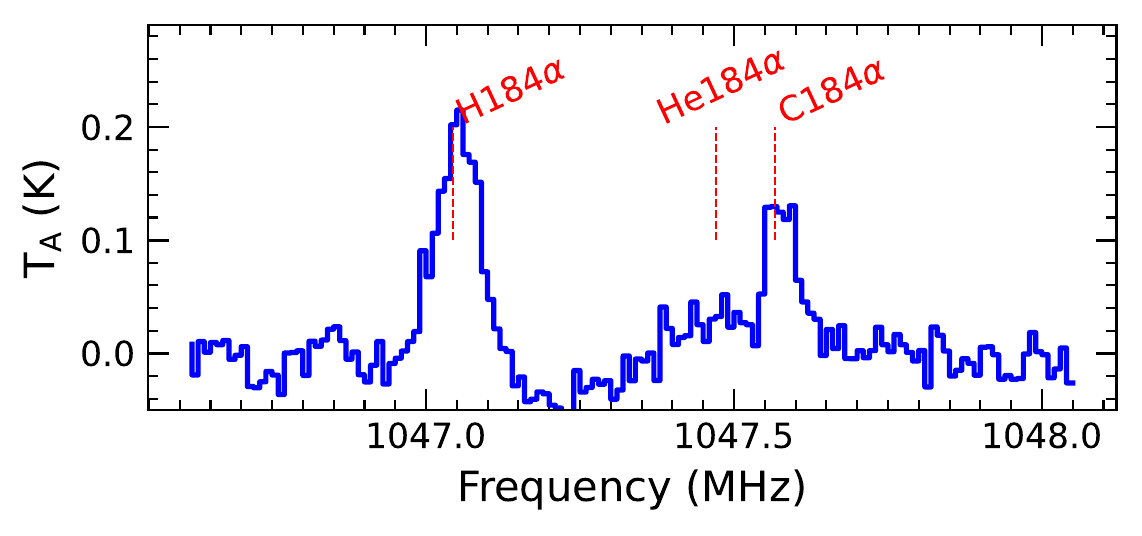}
\includegraphics[width=0.32\textwidth, angle=0]{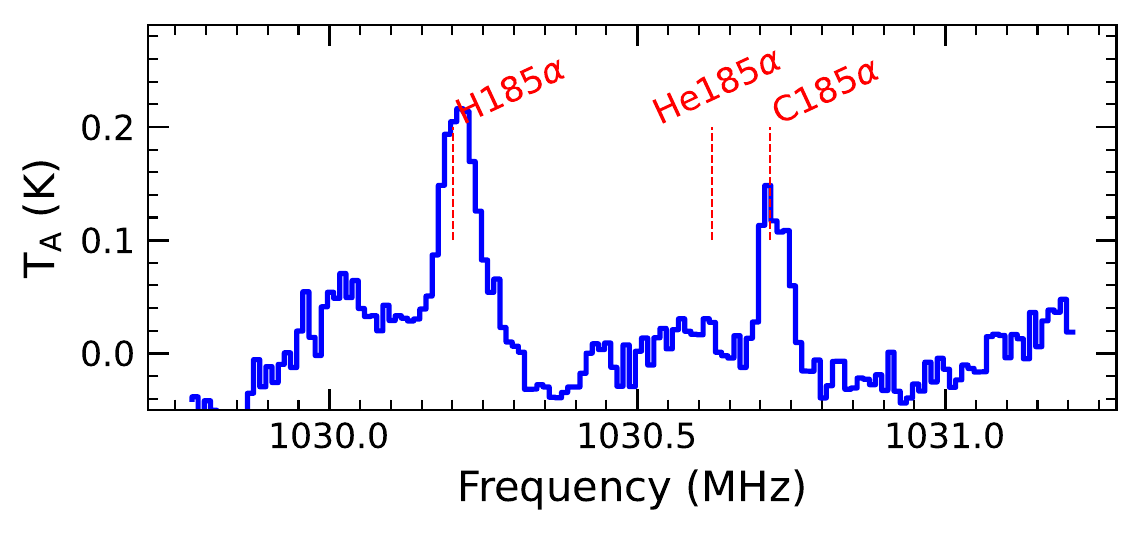}
\includegraphics[width=0.32\textwidth, angle=0]{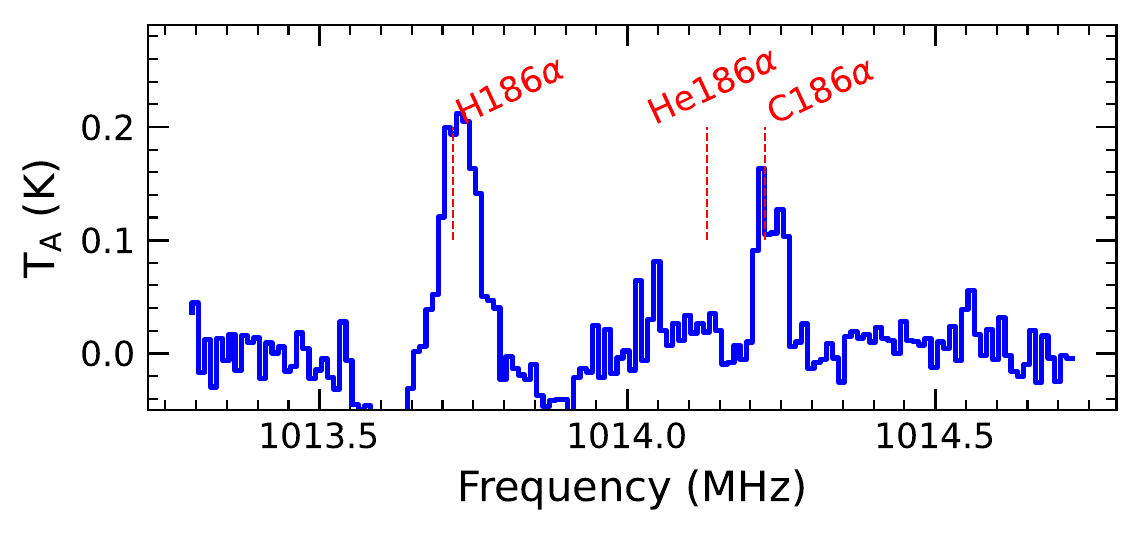}
\caption{Detected RRLs toward G43.148$+$0.013. The positions of H$n\alpha$, He$n\alpha$, and C$n\alpha$ RRLs with $n=164-186$ (except $n=177,180$, which are polluted by RFIs.) have been indicated in each panel.}
\label{Fig:rrl_spectrum}
\end{figure*}

\begin{figure*}
\centering
\includegraphics[width=0.32\textwidth, angle=0]{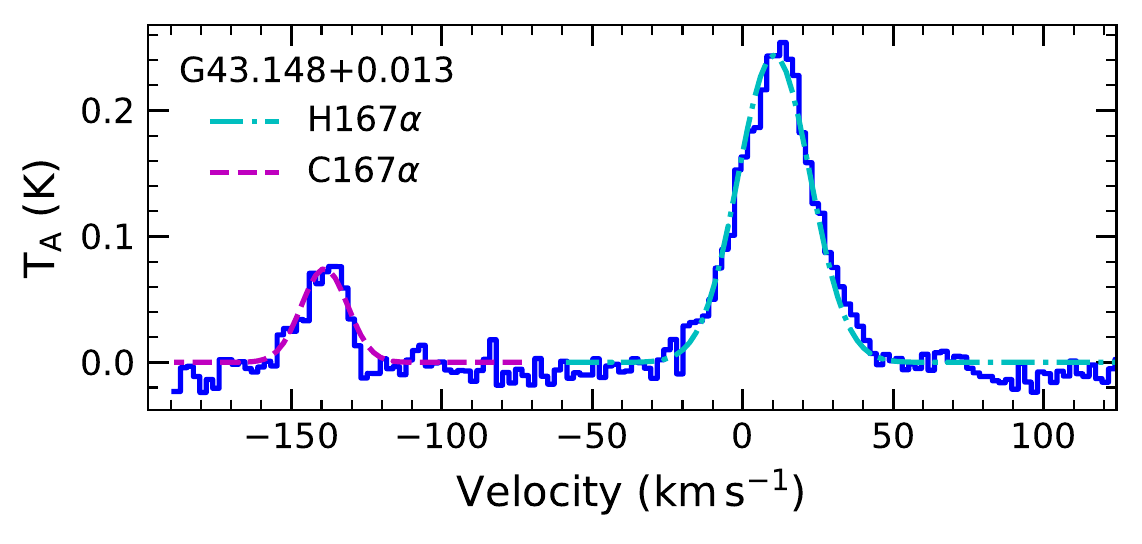}
\includegraphics[width=0.32\textwidth, angle=0]{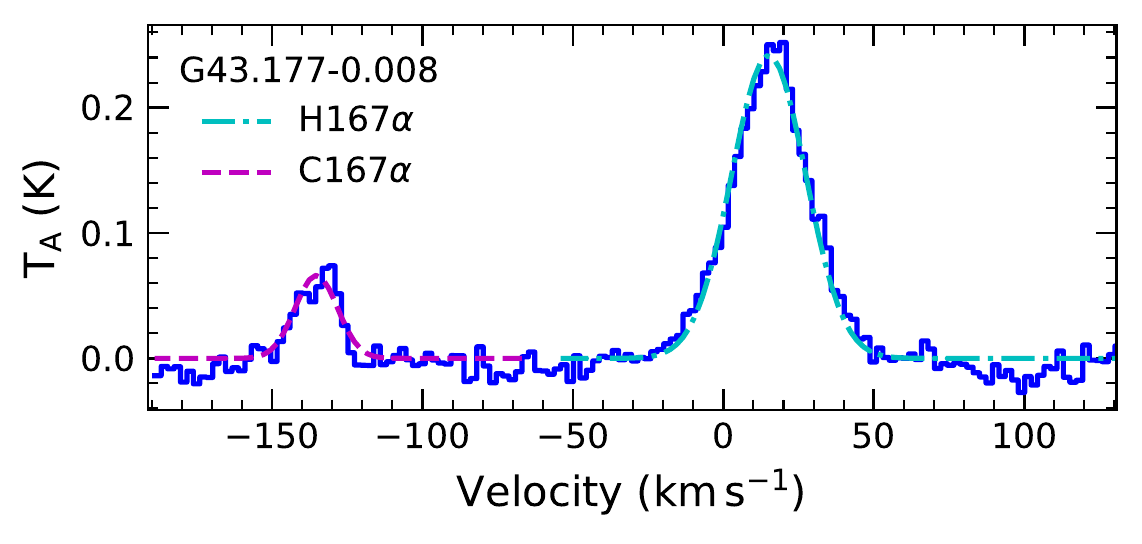}
\includegraphics[width=0.32\textwidth, angle=0]{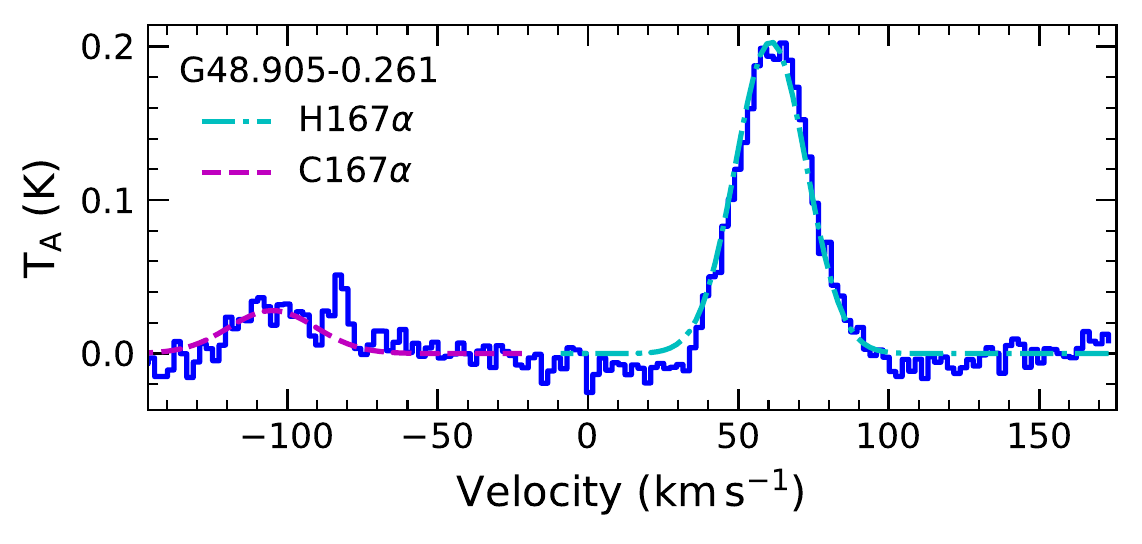}
\includegraphics[width=0.32\textwidth, angle=0]{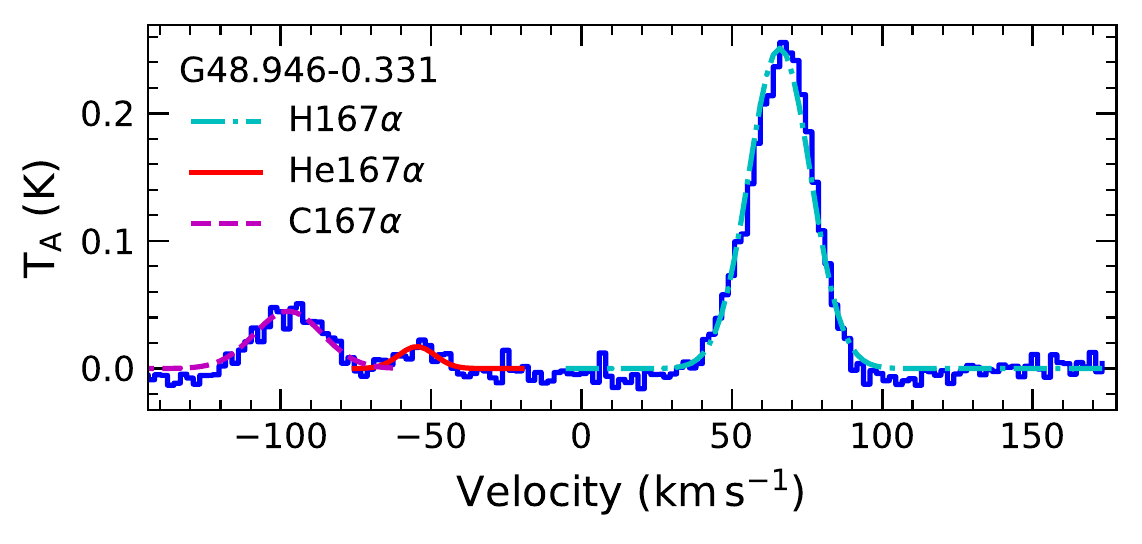}
\includegraphics[width=0.32\textwidth, angle=0]{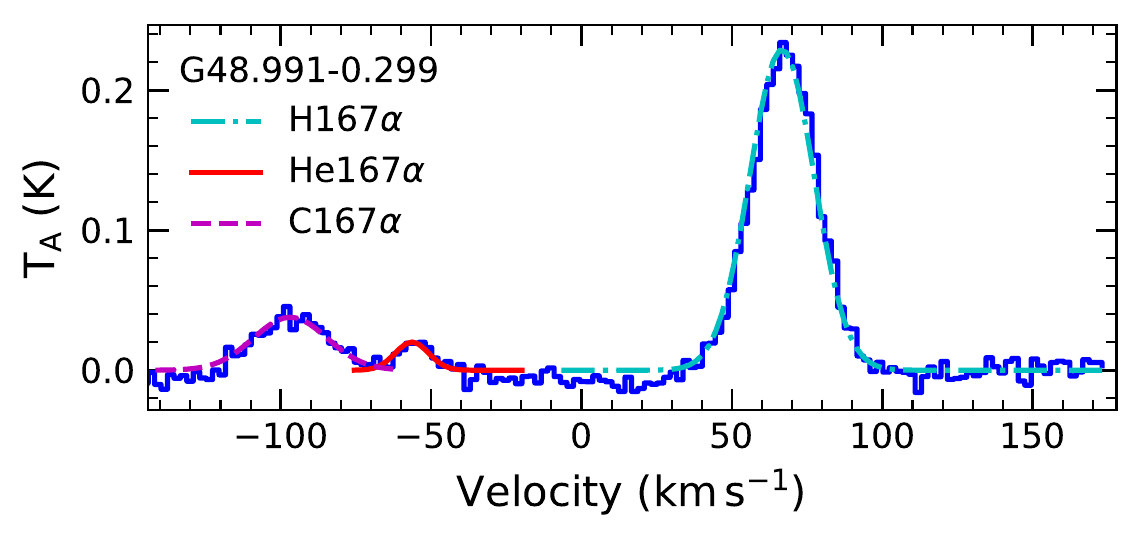}
\includegraphics[width=0.32\textwidth, angle=0]{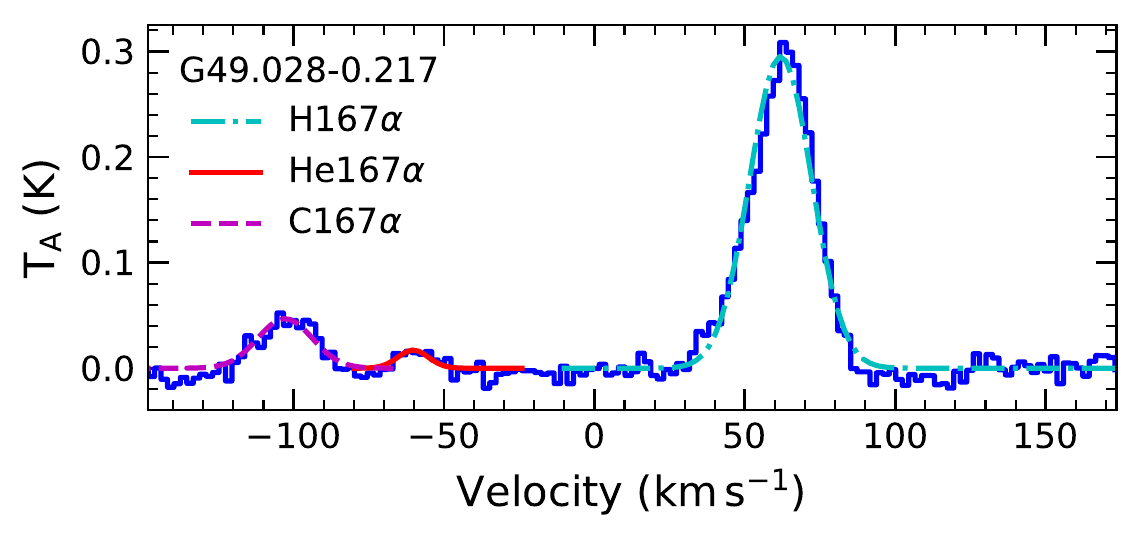}
\includegraphics[width=0.32\textwidth, angle=0]{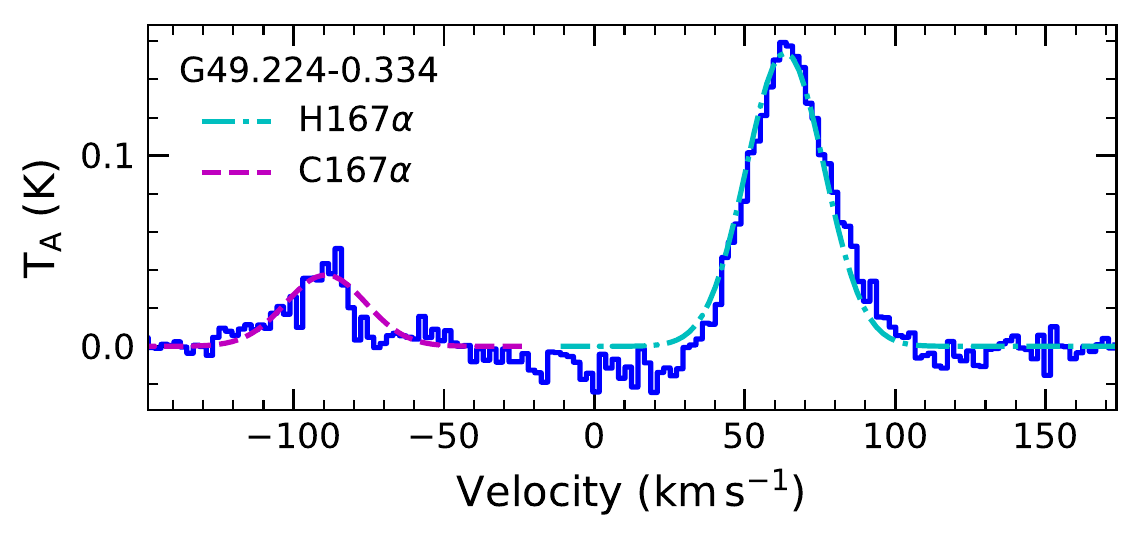}
\includegraphics[width=0.32\textwidth, angle=0]{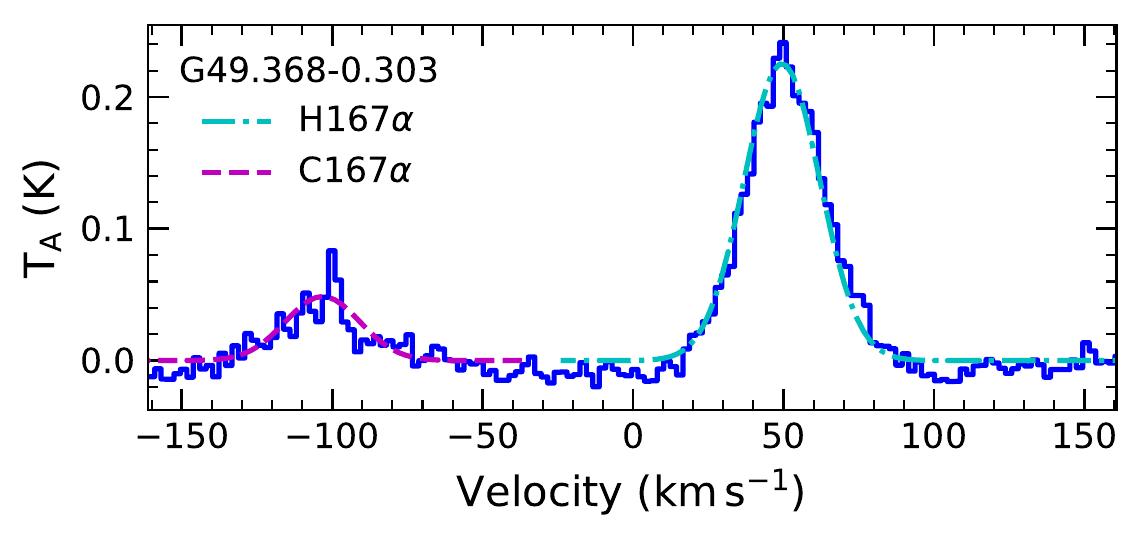}
\includegraphics[width=0.32\textwidth, angle=0]{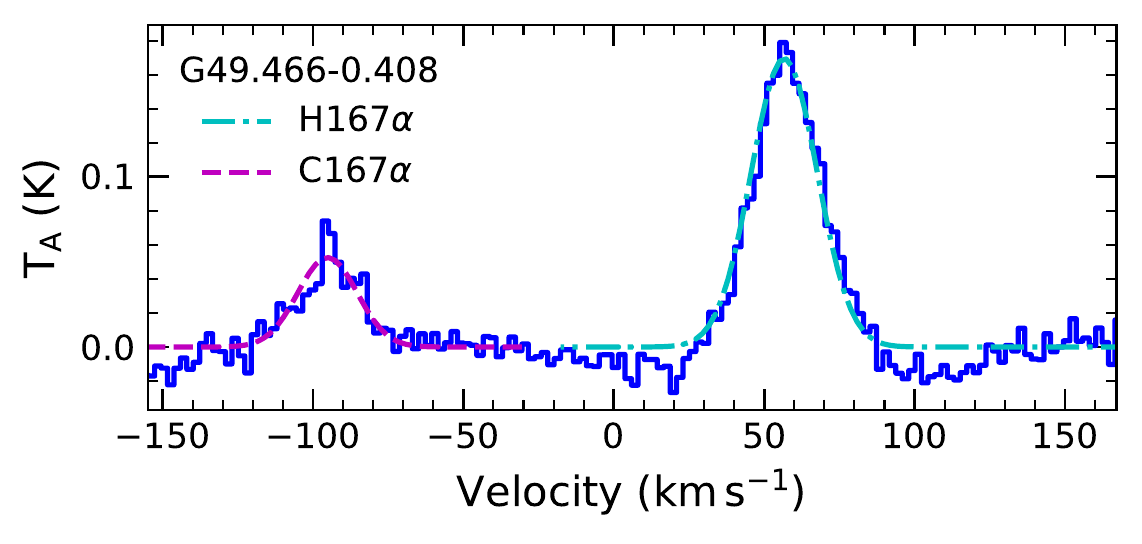}
\caption{Detected H$167\alpha$, He$167\alpha$, and C$167\alpha$ RRLs in nine sources and their Gaussian fitting results. The velocity at 0\,$\kms$ corresponds to 1399.368\,MHz, which is the rest frequency of H$167\alpha$.}
\label{Fig:rrl_velocity}
\end{figure*}

\begin{table}[htp]
%\begin{minipage}[t]{\columnwidth}
\caption{The rest frequencies for H$n\alpha$, He$n\alpha$, and C$n\alpha$ RRLs.}
\label{tab_frequency} 
\centering \small  %\footnotesize %\scriptsize
\setlength{\tabcolsep}{1.6mm}{
\begin{tabular}{cccc}
\hline \hline
Transition &  $\nu_{{\rm H}n\alpha}$&  $\nu_{{\rm He}n\alpha}$&  $\nu_{{\rm C}n\alpha}$  \\
  $n$  &   (MHz)   & (MHz)   & (MHz)     \\  
\hline
164 &  1477.335 &  1477.937  & 1478.072   \\
165 &  1450.716 &  1451.307  & 1451.440   \\
166 &  1424.734 &  1425.314  & 1425.444   \\
167 &  1399.368 &  1399.938  & 1400.066   \\
168 &  1374.600 &  1375.161  & 1375.286   \\
169 &  1350.414 &  1350.964  & 1351.088   \\
170 &  1326.792 &  1327.333  & 1327.454   \\
171 &  1303.718 &  1304.249  & 1304.368   \\
172 &  1281.175 &  1281.697  & 1281.815   \\
173 &  1259.150 &  1259.663  & 1259.778   \\
174 &  1237.626 &  1238.130  & 1238.243   \\
175 &  1216.590 &  1217.086  & 1217.197   \\
176 &  1196.028 &  1196.515  & 1196.625   \\
177 &  1175.927 &  1176.406  & 1176.514   \\
178 &  1156.274 &  1156.745  & 1156.851   \\
179 &  1137.056 &  1137.520  & 1137.624   \\
180 &  1118.262 &  1118.718  & 1118.820   \\
181 &  1099.880 &  1100.328  & 1100.429   \\
182 &  1081.898 &  1082.339  & 1082.438   \\
183 &  1064.307 &  1064.740  & 1064.838   \\
184 &  1047.094 &  1047.521  & 1047.617   \\
185 &  1030.251 &  1030.671  & 1030.765   \\
186 &  1013.767 &  1014.180  & 1014.273   \\
\hline
\end{tabular}}
\begin{flushleft}
\begin{tabular}{l}
%\normalsize
%\textbf{Notes.}
{References: \citet{Muller2005}}
\end{tabular}
\end{flushleft}
\end{table}

\section{Analysis and Discussion}
\label{sect:analysis}

\subsection{RRLs in FAST spectral observations}

Figure\,\ref{Fig:fast_spectrum} presents one FAST spectrum including the whole 500\,MHz bandwidth and using two minutes integration toward \HII region G43.148$+$0.013. H$n\alpha$ ($n=164-186$) RRLs are indicated in the spectrum. Additional bonus are the carbon and helium RRLs, which could be simultaneously covered together with the hydrogen RRLs. This means that the FAST observation could simultaneously cover 23 hydrogen, carbon and helium RRLs, respectively. However, some frequencies are often polluted by RFI. For example at around 1175\,MHz, H177$\alpha$ RRL is always seriously polluted by communication satellites. This will lead to that one often cannot see the seriously polluted emission lines, for example H177$\alpha$ RRL.

\subsection{Hydrogen, Carbon and Helium RRLs}
\label{sect:H_C_He}

Figure\,\ref{Fig:rrl_spectrum} presents the detected RRLs toward the \HII region G43.148$+$0.013. The positions of H$n\alpha$, He$n\alpha$, and C$n\alpha$ RRLs with $n=164-186$ (except $n=177,180$) have been indicated in each panel. For the RRLs with $n=177$ and $n=180$, they have been seriously polluted by RFI, so they are not given here. The rest frequencies of H$n\alpha$, He$n\alpha$, and C$n\alpha$ ($n=164-186$) lines are listed in Table\,\ref{tab_frequency}. Figure\,\ref{Fig:rrl_velocity} shows the detected H$167\alpha$, He$167\alpha$, and C$167\alpha$ RRLs in nine sources (see Table\,\ref{tab_source}) and their Gaussian fitting results. We can see that the He$167\alpha$ RRL was detected in only three sources (G48.946$-$0.331, G48.991$-$0.299, and G49.028$-$0.217), while H$167\alpha$ and C$167\alpha$ RRLs were detected in all the nine sources. The detection rates are respectively 33.3\% for He167$\alpha$ RRL, and 100\% for H167$\alpha$ and C167$\alpha$ RRLs in our selected nine sample.

Information obtained from hydrogen RRLs enables us to determine the basic physical conditions of \HII regions as well as distribution of ionized hydrogen in Galaxy \citep{Brown1978,Zhang2014,Oonk2017,Xu2020}. Helium, the second-most abundance element of the ISM, is also ionized in the majority of \HII regions. The ratio of the integrated intensities of hydrogen and helium RRLs enables us to accurately determine the relative abundance of helium, which has great significance not only for understanding the physics of the ISM, but also for understanding how the Universe formed. In three sources, we simultaneously detected and then measured their line ratios (listed in Table\,\ref{tab_pressure}), which are respectively around 0.0154, 0.0168, and 0.0137 in G48.946$-$0.331, G48.991$-$0.299, and G49.028$-$0.217. This means that in such \HII regions there is $\langle \frac{N({\rm He}^{+})}{N({\rm H}^{+})} \rangle \sim 1.5\%$, which is consistent with the investigations for other giant \HII regions in Galactic center \citep{Churchwell1974}.

\citet{Gordon2002} found that the centroids of the carbon line emission are offset from that of the hydrogen line emission, the radial velocities were often different from the hydrogen RRLs, and the line widths were always much narrower. Although the carbon lines were spatially associated, they probably did not come from the region of the \HII gas itself. For these reasons, \citet{Zuckerman1968} suggested that the carbon lines originated from the outer parts of a dense HI region bounding the discrete \HII regions. Comparing the centroids and line widths of the carbon and hydrogen 167$\alpha$ RRLs for all sample, we found that there exist velocity offsets within $\pm10\,\kms$ between them, and that the carbon line widths were indeed narrower than those for the hydrogen RRLs. This indicates that the \CII and \HII regions are formed in different layers of molecular clouds. For example, \citep{Alves2015} suggested the \CII regions may be formed in the outer and cooler layers of molecular clouds at the boundaries with \HII regions. The helium and hydrogen 167$\alpha$ RRLs have a close line-center velocity, indicating that they may be originated from the same region.

\subsection{Electron density, electron temperature, and pressure}

Interstellar bubbles (\HII regions in nature) surrounding OB stars are the result of the combined effects of radiation pressure and stellar winds \citep{N131,Zhang2016,Zhang2019}. The expansion of bubbles will lead to changes in the dust-to-gas ratio, due to movement of the grains from one zone to another, as well as because of grain destruction \citep{Draine2011}. In the outer part of bubbles, \citet{Draine2011} and \citet{Zhang2013} detected the drift speeds become small as being away from the bubble. Radiation pressure acting on gas and dust causes \HII regions to have central densities that are lower than the density near the ionized boundary \citep{Draine2011,Dutta2018}. Since the effects of radiation pressure are not negligible, the observed cavity must be the result of the combined effects of radiation pressure and a stellar wind.

\begin{table}
\caption{Physical features for three \HII regions.}
\label{tab_pressure} 
\centering \small  %\footnotesize %\scriptsize
\setlength{\tabcolsep}{1.6mm}{
\begin{tabular}{lccccccc}
\hline \hline
Source & Distance & $\frac{N({\rm He}^{+})}{N({\rm H}^{+})}$ & $\frac{T_{\rm 1.4GHz}}{T_{\rm H167\alpha}}$
&  $T_{\rm e}$  & EM & $n_{\rm e}$ & $P_{\rm e}$   \\
     & kpc & & & K & $\rm pc\,cm^{-6}$ &$\rm cm^{-3}$ & Pa   \\  
\hline
G48.946$-$0.331 & $5.3\pm0.2$   &  0.0154 & 81.5 &  7900 & $5.4\times10^4$  &  153 & $3.3\times10^{-11}$ \\
G48.991$-$0.299 & $5.3\pm0.2$   &  0.0168 & 77.9 &  7064 & $4.5\times10^4$  &  139 & $2.7\times10^{-11}$ \\
G49.028$-$0.217 & $5.3\pm0.2$   &  0.0137 & 75.6 &  8986 & $6.1\times10^4$  &  162 & $4.0\times10^{-11}$ \\
\hline
\end{tabular}}
\begin{flushleft}
\begin{tabular}{l}
%\normalsize
%\textbf{Notes.}
{References for distance: \citet{Sato2010}, \citet{Anderson2014}, \citet{Chen2020}}
\end{tabular}
\end{flushleft}
\end{table}

For optically thin ionized gas and local thermodynamic equilibrium (LTE) condition, one could derive the electron temperature $T_{\rm e}$ based on the ratio between 1.4\,GHz continuum flux density and H167$\alpha$ line intensity \citep{Gordon2002,Zhang2014} with
\begin{eqnarray}
T_{\rm e}&=&\Bigg[\left(\frac{6985}{\alpha(\nu,T_{\rm e})}\right)
     \left(\frac{\Delta V_{\rm H167\alpha}}{\rm km~s^{-1}}\right)^{-1}  
     \left(\frac{T_{\rm 1.4GHz}}{T_{\rm H167\alpha}}\right)
    % \times \nonumber \\    &&  
     \left(\frac{\nu}{\rm GHz}\right)^{1.1}
    \left(1+\frac{N({\rm He^+})}{N({\rm H^+})}\right)^{-1}\Bigg]^{0.87},
\end{eqnarray}
where $\alpha(\nu,T_{\rm e}) \sim 1$ is a slowly varying function tabulated by \citet{mezg1967}, $\frac{N({\rm He^+})}{N({\rm H^+})} \sim 0.015$ (see Section\,\ref{sect:H_C_He}), $\Delta V_{\rm H167\alpha}$ is the FWHM of H167$\alpha$ RRL, and the ratio $\frac{T_{\rm 1.4GHz}}{T_{\rm H167\alpha}}$ could be directly measured from the observations, where we ignore the relatively weak background contribution into the 1.4\,GHz continuum. Finally we get the electron temperature $T_{\rm e}$ listed in Table\,\ref{tab_pressure}.

The emission measure (EM) could be derived \citep[e.g.,][]{wils2009,Zhang2014} with
\begin{eqnarray}
{\rm EM} &=& 7.1~{\rm pc~cm^{-6}} 
    \left(\frac{S_{L}}{\rm Jy}\right)
    \left(\frac{\lambda}{\rm mm}\right)
    \left(\frac{T_{\rm e}}{\rm K}\right)^{1.5}  % \times \nonumber \\  &&  
    \left(\frac{\Delta V}{\rm km~s^{-1}}\right)  
    \left(\frac{\theta_{\rm s}}{\rm arcsec}\right)^{-2}. \label{eq:rrlEM}
\end{eqnarray}
For the H167$\alpha$ line, the peak line intensity is $S_L$ and the line width $\Delta V$ are listed in Table\,\ref{tab_source}. The observing wavelength is $\lambda$ = 21\,cm. Based on the parameters above, we could obtain an EM. The intrinsic size $\theta_{\rm s}$ could be derived from the radius of an \HII region. Our selected \HII regions have larger size than the telescope beam ($3'$), thus we could assume $\theta_{\rm s}=1.5'$ for estimation. The corresponding volume electron density is estimated by $n_{\rm e}$ from 
\begin{eqnarray}
{\rm EM} = n_{\rm e}^2Lf_V,
\end{eqnarray}
where we could assume the path length as $L=D\times \rm tan(\theta_{\rm s})$ ($D$ is the source distance listed in Table\,\ref{tab_pressure}) and volume filling factor as $f_{\rm V} = 1$.

The pressure $P_{\rm e}$ from the ionized gas can be estimated following: 
\begin{eqnarray}
P_{\rm e} = 2n_{\rm e}k_{\rm b}T_{\rm e},
\end{eqnarray}
where electron density $n_{\rm e}$ and electron temperature $T_{\rm e}$ have been derived and listed in Table\,\ref{tab_pressure}. If not measured, $T_{\rm e}$ can be inferred from the Galactocentric distance of a source using $R_{\rm gal}$: $T_{\rm e} = 278{\rm\,K} \times(R_{\rm gal}/{\rm kpc}) + 6080 {\rm\,K}$ \citep{Tremblin2014}. Comparing the two methods, the derived electron temperatures in our work are consistent with that. Given a larger sample, we could estimate the mid-plane pressure as a function of Galactocentric distance \citep{Wolfire2003} and test the fitting formula in \citet{Tremblin2014}. This will also help us to further understand the distribution of $n_{\rm e}$ and $T_{\rm e}$ in the Milky Way.

\section{Summary}
\label{sect:summary}

The FAST telescope has a large collecting area and enables high-sensitivity observations for extended sources. The 19-beam receiver has a wide bandwidth (500\,MHz) and it can simultaneously cover 23 H$n\alpha$, He$n\alpha$, and C$n\alpha$ ($n=164-186$) RRLs, respectively. This combination is idea for the systematical study of RRLs from Galactic \HII regions. 
 In this pilot study, we observe nine \HII regions using  hydrogen, carbon and helium RRLs. Using 20\,minutes pointing time for each source, a sensitivity of around 9\,mK with a spectral resolution of around 2.0\,$\kms$ can be achieved. In total, 21 RRLs for H$n\alpha$ and C$n\alpha$ at $1.0-1.5$\,GHz have been simultaneously detected in strong emission except two lines ($n=177$ and $n=180$), which are seriously polluted by RFI. Overall, the detection rates for the He$167\alpha$ and C$167\alpha$ RRLs are 100\%, while that for the He$167\alpha$ RRL is 33.3\% in our selected sample.

Comparing the centroids and line widths of the carbon and hydrogen 167$\alpha$ RRLs for all sample, we found that there exist velocity offsets between $-10$ and $+10\,\kms$. This indicates that the \CII and \HII regions are formed in different layers of molecular clouds. Using the hydrogen and helium RRLs, we have measured electron density, electron temperature, and pressure (listed in Table\,\ref{tab_pressure}) in three \HII regions. This pilot study proves that the survey for a large sample of statistical measurement is feasible in the future.

In a future FAST work, we plan to observe a statistically significant sample of \HII regions to measure electron density, electron temperature, pressure, evolutionary stages, and the relative abundance of elements. The sample would allow us to study the distribution of these physical parameters (including line widths, line ratios, electron temperatures and densities) in various parameter spaces,  as well as the relation between the physical properties of \HII and their locations measured in terms of, for example, the Galactocentric distance, to understand the co-evolution between the \HII regions and the Galactic disk. 

\section*{Acknowledgements}
% Entry for the table of contents, for this guide only
\addcontentsline{toc}{section}{Acknowledgements}

%We wish to thank the anonymous referee for comments and suggestions that improved the clarity of the paper.
C.P.Z acknowledges support by the NAOC Nebula Talents Program and the Cultivation Project for FAST Scientific Payoff and Research Achievement of CAMS-CAS. J.L.X, L.G.H, and N.P.Y thank the support from the Youth Innovation Promotion Association CAS. This work is supported by the National Natural Science Foundation of China Nos.\,11703040, W820301904, 11988101, 11933011, and 11833009. This work has used the data from the Five-hundred-meter Aperture Spherical radio Telescope (FAST). FAST is a Chinese national mega-science facility, operated by the National Astronomical Observatories of Chinese Academy of Sciences (NAOC).

\bibliographystyle{raa}
\bibliography{references}

\end{document}